%% file: jhy-p1-21-is_twocol.tex
\documentclass[journal]{IEEEtran}

\usepackage{amsmath}
\usepackage{mathtools,nccmath}
\usepackage{amsfonts,dsfont}
\usepackage{MnSymbol}
\usepackage{color}
\usepackage{hyperref}
\usepackage[noadjust]{cite}
\usepackage[linesnumbered, ruled, lined]{algorithm2e}
\usepackage{graphicx}
\usepackage[caption=false]{subfig}

\def\be{ \begin{eqnarray} }
\def\ee{ \end{eqnarray} }

\begin{document}

\title{
Design of Anti-Jamming Waveforms for\\
Time-Hopping Spread Spectrum 
Systems in\\
Tone Jamming Environments}

\author{Hyoyoung~Jung, Binh~Van~Nguyen, Iickho~Song, and Kiseon~Kim

\thanks{Copyright (c) 2015 IEEE. Personal use of this material is permitted. However, permission to use this material for any other purposes must be obtained from the IEEE by sending a request to pubs-permissions@ieee.org.}
\thanks{ 
	H.~Jung and K.~Kim are with the School of
	Electrical Engineering and Computer Science,
	Gwangju Institute of Science and Technology, Gwangju 61005
	Republic of Korea (e-mail: \{rain, kskim\}@gist.ac.kr).}
\thanks{B.~V.~Nguyen is with the Institute of Research and Development, Duy Tan University,
Da Nang 550000 Vietnam (e-mail: nguyenvanbinh12@duytan.edu.vn and n.v.binh1986@gmail.com).}
\thanks{I.~Song is with the School of Electrical Engineering, Korea Advanced Institute of
Science	and Technology, Daejeon 34141 Republic of Korea, and also with Liangjiang
International College, Chongqing University of Technology, Chongqing 401135
People's Republic of China (e-mail: i.song@ieee.org).}
\thanks{The authors gratefully acknowledge the support from the Electronic Warfare
	Research Center 
	at Gwangju Institute of Science and Technology, 
	originally funded by the Defense Acquisition Program Administration and 
	Agency for Defense Development, 
	and the support from the National Research Foundation of Korea
	under Grant NRF-2018R1A2A1A05023192.}
}

\markboth{Accepted to IEEE Transactions on Vehicular Technology}%
{Jung \MakeLowercase{\textit{et al.}}: IEEEtran.cls for IEEE Journals}

\maketitle
\begin{abstract}
We consider the problem of designing waveforms for mitigating single tone
jamming (STJ) signals with an estimated jamming frequency in time-hopping  
spread spectrum (TH SS) systems. The proposed design of waveforms optimizes
the anti-jamming (AJ) performance of TH SS systems
by minimizing the correlation 
between the template 
and STJ signals, in which the problem of waveform optimization is simplified
by employing 
a finite number of rectangular pulses. The simplification
eventually makes the design of waveforms be converted into a problem of
finding eigenvalues and eigenvectors of a matrix.
Simulation results show that the waveforms designed by the proposed
scheme 
provide us with 
performance superior not only to the conventional waveforms but also
to the clipper receiver in the mitigation of STJ.
The waveforms from the proposed design
also 
exhibit a desirable AJ
capability 
even when the estimated frequency of the STJ is not perfect.

\end{abstract}


\section{Introduction}
%
%

\IEEEPARstart{T}{he}
spread spectrum (SS) techniques, spreading the
bandwidth of a signal beyond that 
actually required, 
have been developed not only for commercial communication but also
for covert communication.
%
Systems adopting SS techniques have been effectively utilized for the multiple
access capability, suppression of interference, alleviation of multipath fading
effects, and resilience against jamming signals \cite{Poisel}.
In particular, among the three classes
direct sequence (DS), frequency-hopping (FH),
and time-hopping (TH) of SS systems, the TH SS systems,
modulating the transmission signal by shifting it arbitrarily in
time, have been widely used due to their
better (compared to the DS and FH SS systems)
capability of resolving multipath, easiness in implementation
with low complexity \cite{THUWB-00},
and low probability of interception \cite{Poisel}.

In addition to pulse position modulation (PPM), many pulse amplitude modulation
schemes such as the binary phase shift keying (BPSK) and on-off keying 
\cite{PPM-BPSK-Anal} can be incorporated in the TH SS systems.
In \cite{PPM-BPSK-Anal,NBI-MPC-Anal,BPSK-MUI-Anal}, the bit error rate (BER)
performance of TH SS systems with various modulation schemes has been
analyzed in channels with additive white Gaussian noise (AWGN), multipath
fading channels, and multiple access interference (MAI).
In \cite{HIHO-capa-anal}, the hard-input-hard-output capacity of TH-BPSK
system is analyzed in the presence of timing error with interpath,
interchip, and intersymbol interferences.
In \cite{MAI-Whi-Rx}, the MAI-plus-noise whitening filter was proposed to
improve the performance of a TH-BPSK system.

In the meantime, as the hardware and software technology advances,
modern jamming attacks become more complex and intelligent
from single tone jamming (STJ), a special case of tone jamming (TJ),
into
the multi tone jamming (MTJ), time-varying STJ (TV-STJ),
and sweep jamming (SWJ), for instance \cite{SWJ,PAC-US-AJ,TJ,WCM-GameAJ}.
Nonetheless, the TJ (especially the STJ) is commonly considered and assumed
in the investigation of anti-jamming (AJ) schemes for TH SS systems: This is because
of the fact that the bandwidths of the TH SS systems are far wider than those of conventional
systems \cite{NBI-MPC-Anal} and that most jammers would choose the STJ for its simplicity,
effectiveness, and jamming efficiency
with a concentrated power on the channel \cite{THseq-Design, THseq-Design-21}.
%

Obviously,
analyzing and improving the anti-TJ performance of TH SS systems are of
paramount importance and have been
investigated in various studies. 
For instance,
the BER performance of a TH SS system is analyzed under multipath
fading and TJ environment in \cite{NBI-MPC-Anal}, and
design of TH sequence and pulse shaping for a TH SS system
in TJ environment is addressed in \cite{THseq-Design}.
In \cite{Notch-Filter} and \cite{Para-Selec}, designing a notch filter and
tuning of system parameters, respectively,
for mitigating the effects of TJ on TH SS systems
have been discussed. 
Let us also note that
the joint optimization of
power allocation schemes with the channel selection of
SS systems and user scheduling are proposed for
the AJ purpose under some intelligent jamming scenarios in \cite{PAC-US-AJ}
and \cite{FH-PAC}.

%
%
%
%
%

Analyses of the AJ performance of TH SS systems
and designing TH pulses for spectral efficiency
\cite{Pulse-Design-07}, multiple access performance \cite{Pulse-Design-08},
compliance
with spectral emission constraints \cite{Pulse-Design-06,Pulse-Design-17} have been
carried out extensively. Yet, it seems that design of waveforms for the purpose of AJ
has not attracted much interest so far.
In this paper,
to improve the anti-TJ performance of TH SS systems at the transmitter side,
we propose
an optimal design of waveforms against STJ.
In the proposed design of
waveforms,
the AJ performance of TH SS systems is optimized by minimizing the correlation
between
the template signal of the TH systems and the estimated jamming signal.
The problem of designing waveforms is first simplified
by approximating the 
continuous waveform to be designed
via a linear combination
of a finite number of rectangular pulses.
After some further manipulations,
we have shown that
%
%
the simplified problem can be solved with optimization
techniques such as Powell's conjugate--direction method.

Simulation results
confirm that the waveforms 
designed by the proposed scheme 
enhance the AJ performance of TH SS systems in the TJ environment
even when the estimation of the jamming frequency is
imperfect. 
%
%
It is noteworthy that
design of waveforms is also one of the key elements 
of the waveform reconfiguration in game-theoretic strategies
for developing intelligent AJ communication systems
that react against the counterpart in a timely manner \cite{WCM-GameAJ,TVT-GameAJ}.

The main contributions of this paper can be summarized 
as follows:
	\begin{itemize}
		\item An optimal design of waveforms for TH SS systems against STJ is proposed
for the improvement of anti-TJ performance at the transmitter side.
		\item The problem of designing waveforms is simplified by waveform approximations
and analytic derivations. The simplified problem can be solved by finding the eigenvectors
of a matrix, for which many common optimization techniques can be employed.
		\item The waveforms designed by the proposed scheme outperform the conventional
waveforms of TH SS systems in the STJ environment for both the ideal and imperfect
estimation of the jamming frequency.
		\item The proposed design of waveforms could be employed as a key function of
waveform reconfiguration for developing game-theoretic AJ communication systems.
	\end{itemize}

The remainder of the paper is organized as follows.
In Section \ref{sec-model}, we describe 
the system and jamming models together
with some preliminary notions in the proposed design of waveforms for AJ purposes.
Section \ref{sec-design} is devoted to the description of the proposed procedure
of designing waveforms, followed by discussions on numerical and simulation results
in Section \ref{sec-simul}. Section \ref{sec-concl} summarizes this paper.

\section{System 
and Jamming Models}\label{sec-model}

\subsection{TH SS System Model}

\begin{figure}[t!]
	\centering
	\includegraphics[width = 8.5cm]{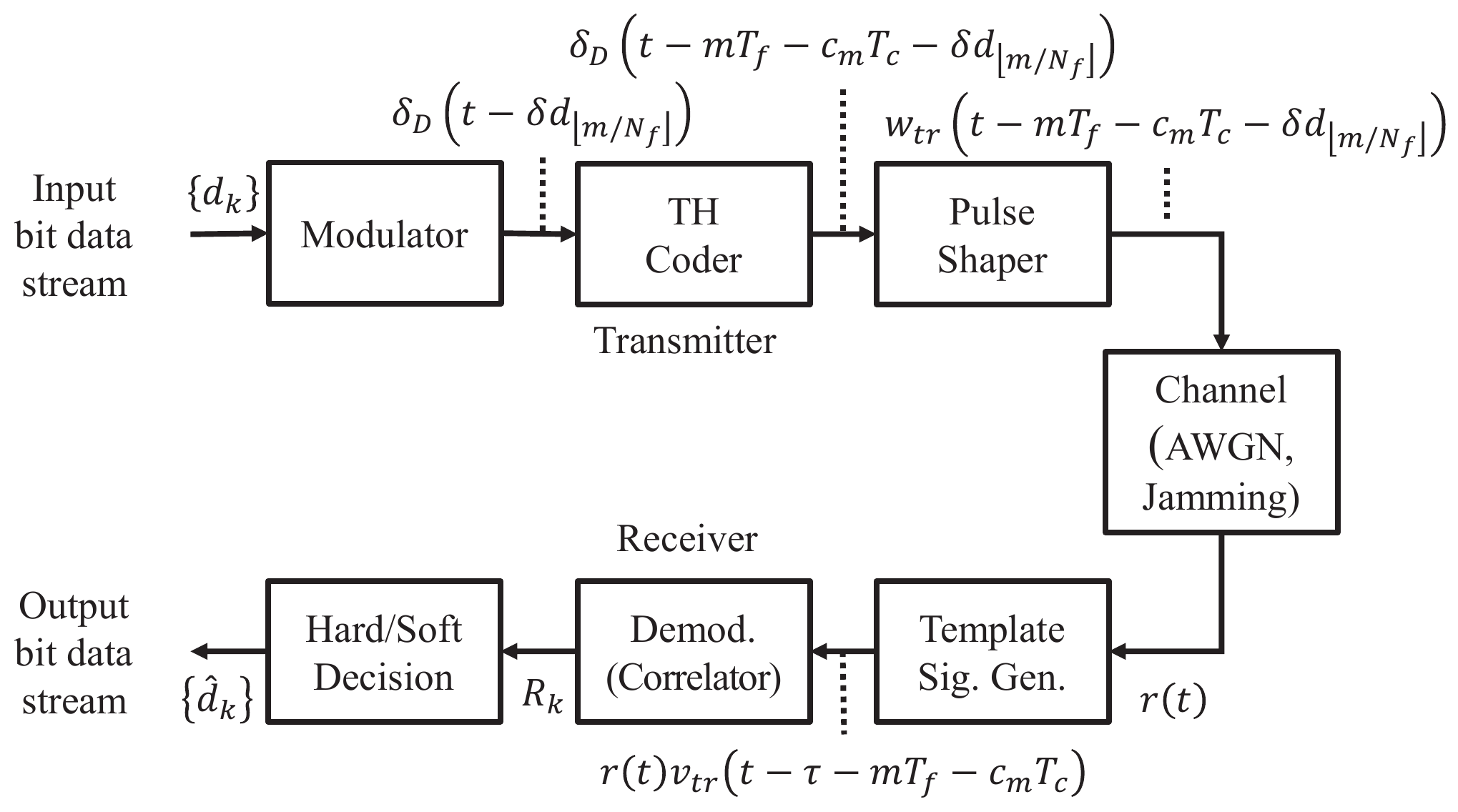} 
	\caption{Block diagram of TH systems.}\label{fig:th_system}
\end{figure}

A block diagram of TH systems is shown graphically in Fig. \ref{fig:th_system}.
In this figure,
$ \delta_D(t) $ is the Dirac delta function \cite{Dirac-Delta} and
$ \left\lbrace d_k\right\rbrace  $ denotes 
a stream of data bits, in which each bit is of duration
$ T_b $ 
composed of $ N_f=T_b/T_f $ frames of duration $ T_f $.
A frame in turn consists of $ N_c=T_f/T_c $ chips of duration $ T_c $, where
$ T_c \ll T_f $ in practice.
The transmitted waveform $w_{tr}(t)$ of the TH system, in the form of pulses
with a very short duration, is often called 
a monocycle.
The duration $ T_p $ of a monocycle is usually chosen to be $ T_p < T_c $ and
is in the order of $1$ ns yielding a bandwidth in the order of $1$ GHz.

For the transmission of the data stream
$\left\lbrace d_k\right\rbrace $, we first generate a TH
code $ \left\lbrace
c_m \right\rbrace_{m=1}^{N_h} $, a set of 
independent and identically distributed  
random variables 
over $ \left\lbrace 1, 2, ..., N_c\right\rbrace$.
The TH code determines the location of a chip within a frame:
For example, when $ c_5=3 $, we transmit a monocycle at the third chip in
the fifth frame.
%
When the TH-PPM is employed,
a monocycle is delayed by the PPM shift $\delta$ and $0$ for
a data bit `1' and `0', respectively.
The transmitted waveform of the TH-PPM can then be written as \cite{THUWB-00}
\begin{equation}\label{eq_tx_TH_PPM}
s_{PPM}(t)=\sum_{m=-\infty}^{\infty} w_{tr}\left
 ( t-mT_f-c_mT_c-\delta d_{\lfloor m/N_f\rfloor}\right ),
\end{equation}
where $\lfloor \cdot \rfloor$ is the floor function.
%
%
The signal $r(t)$ received at the receiver can subsequently be written as
\begin{equation}\label{eq_rx_model}
r(t)=\alpha s_{PPM}(t-\tau)+j(t)+n(t),
\end{equation}
where $ \alpha $ is the channel gain, $ \tau $ is a random variable over
$ \left[ 0, \infty \right)  $
representing the time asynchronism between the transmitter and receiver,
$j(t)$ is the jamming signal, and $n(t)$ is the AWGN.

When a correlator receiver is employed,
the received signal $ r(t) $ is correlated with the template signal
\be
v_{tr}(t)=w_{tr}(t)-w_{tr}(t-\delta)
\ee
of the monocycle $w_{tr}(t)$.
Assuming a perfect synchronization ({\it i.e.}, $\tau$, $c_m$, $T_f$, and $T_c$
are available at the receiver), the output of the correlator for the $k$-th
bit can be expressed as
\be\label{eq_re_corr}
R_k &=& 
\sum_{m=kN_f}^{(k+1)N_f-1}
\int_{\tau+mT_f}^{\tau+(m+1)T_f}
r(t)
\nonumber \\
&& \quad \times v_{tr}\left (t-\tau-mT_f-c_mT_c \right) dt 
\nonumber \\ 
&=&S_k + J_k + N_k,
\ee 
where $S_k, J_k$, and $N_k$ are the correlator outputs corresponding to the TH
signal $s_{PPM}(t)$, jamming signal $j(t)$, and AWGN $ n(t) $, respectively.
The decision hypothesis that a data bit of `0' is sent is chosen if
$R_k \ge 0$: Specifically,
\be
\widehat{d}_k = 
\frac{1}{2}\left \{1-\mbox{sgn}\left (R_k \right )\right \}
\ee
is the estimate of $d_k$.

\subsection{Tone Jamming Models}
When a jamming signal has a single and a multitude of frequencies,
it is called an STJ 
and a MTJ signals, respectively \cite{TJ}.
In this paper, we assume the STJ model, which can be a serious attack with a
concentrated power.
The jamming signal $ j(t) $ in \eqref{eq_rx_model} can
be expressed more specifically as
\begin{equation}\label{eq:j_STJ}
j_{STJ}\left ( t;f_J, \theta_J \right )=\sqrt{2P_J} \cos \left (2\pi f_J t
+\theta_J \right ),
\end{equation}
where $P_J$, $f_J$, and $\theta_J$ are
the power, frequency, and phase, respectively, of the STJ signal.

\subsection{Correlation between TH and Jamming Signals}

Let us denote the
correlation function 
between the jamming and template signals
by
%
\begin{equation}\label{eq_R_tr_STJ}
\begin{aligned}
&R_{tr,T}\left (z;\theta_J, f_J, T_c \right )\\
&=\int_{0}^{T_c}{v_{tr}(t) \cos(2\pi f_J \left (t+z)+\theta_J \right ) dt}
\end{aligned}
\end{equation}
using \eqref{eq:j_STJ} for an
STJ, where the constant $\sqrt{2 P_J}$ is omitted
for simplicity.

Since the template signal $v_{tr}\left ( t-\tau -mT_f-c_mT_c \right )$
has non-zero
values only during one chip duration of $T_c$,
{\it i.e.},
\be
mT_f+c_mT_c \le t-\tau\le mT_f+(c_m+1)T_c
\ee
per one frame, the correlator output $ J_k $  in
\eqref{eq_re_corr}
of the jamming signal
for the $ k $-th bit can be expressed in general as
\be\label{eq_JktoR_tr_j}
J_k &=& \sum_{m=kN_f}^{(k+1)N_f-1} \int_{\tau+mT_f}^{\tau+(m+1)T_f}
j(t) v_{tr}\left (t-\tau-mT_f-c_mT_c \right) dt \nonumber \\
&=& \sum_{m=kN_f}^{(k+1)N_f-1}
\int_{\tau+mT_f+c_m T_c}^{\tau+mT_f+ \left (c_m+1\right) T_c}
j(t) \nonumber \\
& & \quad \qquad \qquad \times v_{tr}\left (t-\tau-mT_f-c_mT_c \right) dt \nonumber \\
%
%
&=&\sum_{m=kN_f}^{(k+1)N_f-1}{R_{tr,T}\left ( \Delta_m; \theta_J, f_J, T_c \right )},
\ee
where 
\be
\Delta_m = \tau + mT_f + c_mT_c
\ee
denotes the time shift between the template signal
$v_{tr}\left (t-\tau-mT_f-c_mT_c \right)$ of the TH system and
jamming signal $j(t)$ at the $ m $-th frame.
It is noteworthy that possible values of the time shift 
$\Delta_m$ in (\ref{eq_JktoR_tr_j}) under the influence of the
STJ signal \eqref{eq:j_STJ} are
\be
\mathrm{mod}\left(\mathrm{mod}\left(n T_J-\tau, T_c \right ),T_J\right),
\ee
where $n$ is an integer and $T_J=1/f_J$ is the period of the jamming signal.
Since $ \tau $ is a random variable and $T_c$ is not an integer multiple of
$T_J$ in general, $\Delta_m$ is also a random variable.

\subsection{Clipper}

\begin{figure}[t!]
	\centering
	\includegraphics[width = 8.5cm]{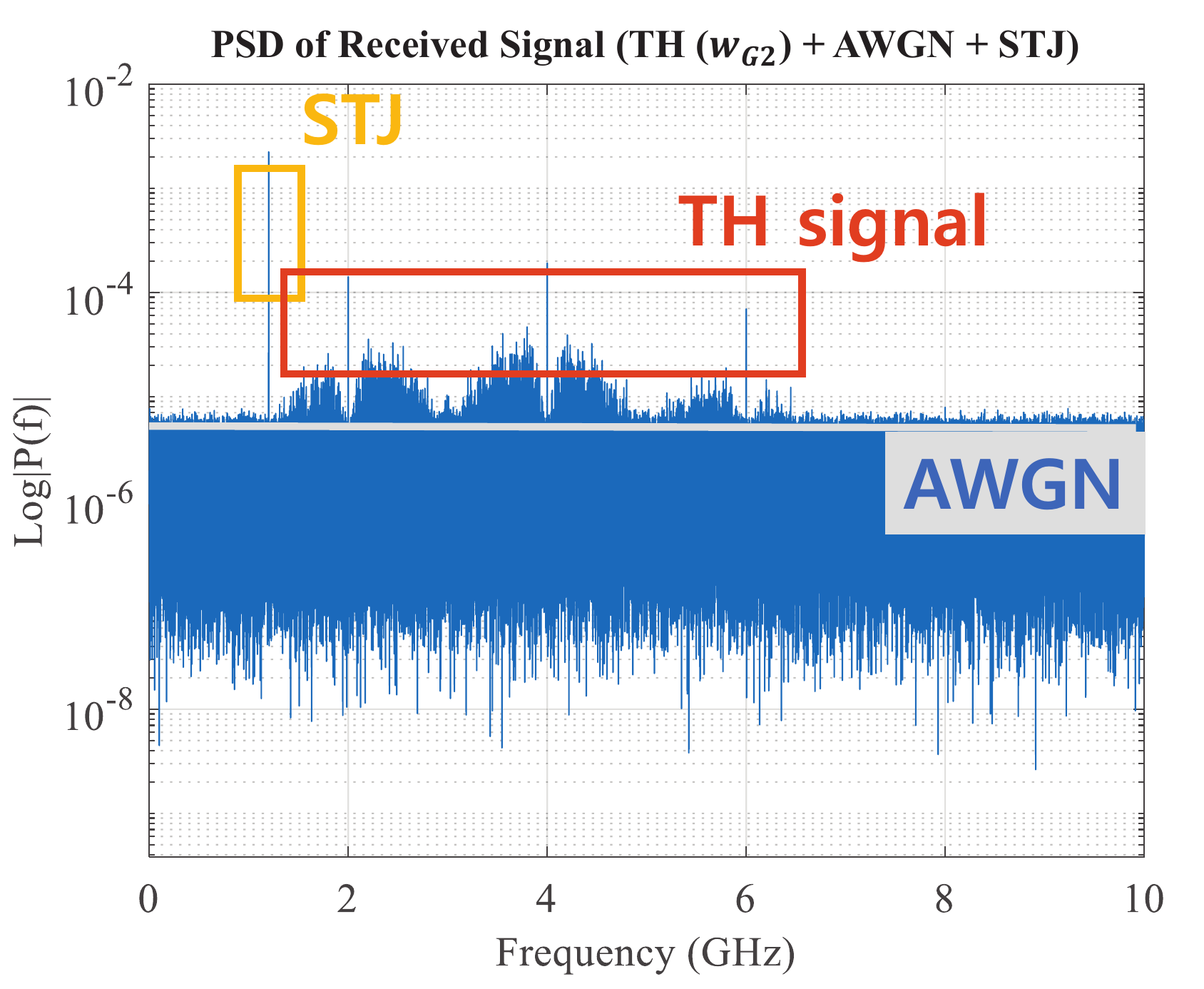} 
	\caption{PSD of the received signal with the components of
TH signal, STJ signal, and noise specified.}\label{fig:SJN_PSD}
\end{figure}


Fig. \ref{fig:SJN_PSD} shows an example of the power spectral density (PSD)
$P_{rec}(f)$ of the received signal $r(t)$ under the STJ model
 in terms of the TH, STJ, and AWGN components when the Gaussian doublet
%
%
\be
 w_{G2} \left ( t+T_m \right ) = A\left\{ 1-4\pi\left (\frac{2 t}{T_p}
  \right )^2\right\}
\exp \left\{ -2\pi\left (\frac{2 t}{T_p} \right )^2 \right\},
\ee
one of the waveforms employed most commonly in TH systems,
is used as the monocycle, where
%
%
$ A $ is the amplitude and $ T_m $ denotes the center 
of the waveform.

The clipper \cite{Clipper},
a simple AJ filter with 
a threshold,
limits the signal based on the second largest value of $\left | P_{rec}(f)
\right |$: Normally,
an STJ signal exhibits the maximum peak of the PSD
due to its concentration at a frequency, and
the second largest value is $\max|P_{TH}(f)|$, 
where $ P_{TH}(f) $ is the PSD of the TH signal.
Therefore, 
the clipping threshold $\lambda_C$ is  
determined as
\begin{equation}\label{eq:clipper}
\lambda_C = K \max|P_{TH}(f)|,
\end{equation}
where the constant $K$ is often 
selected in the interval
$ [1,1.5] $: In this paper,
we choose $K=1.2$, and
assume a clipper at the receiver 
for all the waveforms
later in the comparisons and further investigation of
the AJ performance of the waveform from the proposed design.

\section{Design of Waveforms} 
\label{sec-design}

\subsection{Problem Formulation}
We now consider the
design of waveforms to enhance the AJ capability of the TH system against the STJ.
The 
correlation function $ R_{tr,T} $ 
in \eqref{eq_JktoR_tr_j}
is a sinusoidal function of $\Delta_m$.
Since the maximum value of the correlator output with respect to
$\Delta_m$ determines the AJ performance,
the cost function is
%
\be
\max_{\Delta_m} \left|R_{tr,T}\left
(\Delta_m;\theta_J, f_J, T_c \right )\right| .
\ee

Let us try to find the optimal waveform $\widehat{w}_{tr}(t)$ that minimizes
the cost function under the constraint $\left\| \widehat{w}_{tr}(t)\right\|^2=1$.
We assume that $ T_c $ is given and 
$\theta_J$ is equal to zero, and thus the term $\theta_J$ will not be shown
explicitly 
from now on.
Then, denoting by
$ \widehat{f}_J $
the estimate of the actual jammer frequency $f_J$, 
the problem of waveform design can be formulated as
\begin{equation}\label{eq_Opt_knwonF}
\begin{aligned}
\widehat{w}_{tr,\widehat{f}_J}(t)=&\arg \min_{w_{tr}} \max_{\Delta_m} \left|R_{tr,T}
\left (\Delta_m; \widehat{f}_J, T_c \right )\right|\\
&\mbox{such that }\left\| \widehat{w}_{tr}(t)\right\|^2=1.
\end{aligned}
\end{equation}


\subsection{Problem Simplification}

\begin{figure}[t!]
	\centering
	\includegraphics[width = 8.5cm]{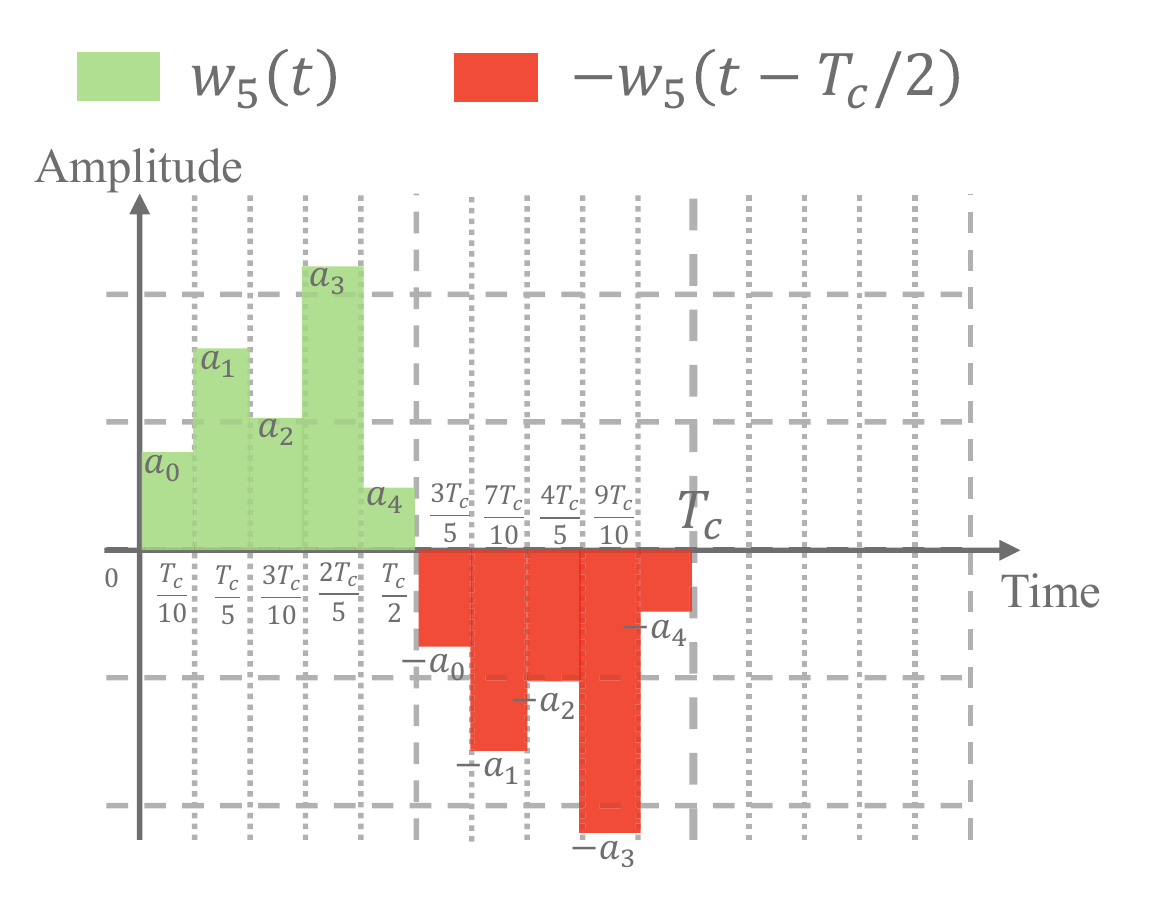} 
	\caption{Approximation of waveform $ w_{tr}(t) $ as a weighted sum of
$ N=5 $ rectangular pulses}\label{fig:w5simp}
\end{figure}

The problem \eqref{eq_Opt_knwonF} is rather intractable mainly due to
the 
minimization and maximization over
continuous spaces 
involved in the optimization.
To make the problem somewhat tractable, we consider an approximation
\begin{equation}\label{eq_w_N}
w_N(t)=\sum_{i=1}^{N}{a_i \mbox{rect}\left( \frac{t-(2i-1)\tfrac{T_c}{4N}}
{\tfrac{T_c}{2N}}\right)}
\end{equation}
of the TH waveform 
$\widehat{w}_{tr,\widehat{f}_J}(t)$,
where $\left\{ a_i \right\}_{i=1}^{N}$ are the weights for
the $ N $ rectangular pulses $ \mbox{rect}(\cdot) $ of duration $ {T_p}/{N} $
with $ T_p=T_c/2 $ as shown in Fig. \ref{fig:w5simp}.
Then, with the template signal
\be
v_N(t)=w_N(t)-w_N\left ( t- \frac{T_c}{2} \right ),
\ee
the 
correlation $ R_{tr,T} $ in \eqref{eq_JktoR_tr_j} 
%
can be expressed as
\be\label{eq_R_N_STJ}
&&R_{N,T}\left (\Delta_m;f_J, T_c \right )
\nonumber \\
&&  \quad = A_N \left (f_J;T_c \right ) B_N \left (
\Delta_m,f_J,\left\lbrace a_i\right\rbrace ;T_c \right )
\ee
by substituting \eqref{eq_w_N} and $\theta_J=0$ in \eqref{eq_R_tr_STJ}, where
%
%
\begin{equation}\label{eq_AN}
A_N \left (f;t \right )= \frac{2}{\pi f}\sin\frac{\pi f t}{2}
\sin\frac{\pi f t}{2N}
\end{equation}
and
\begin{equation}\label{eq_BN}
%
B_N\left (\alpha,f,\left\lbrace a_i\right\rbrace ;t \right )=
\sum_{i=1}^{N}{a_i
\sin \left \{ \pi f \left(\frac{N-1+2i}{2N}t+2\alpha\right) \right \}}. 
\end{equation}

By noting that
$A_N \left (f_J;T_c \right )$ is independent of $\Delta_m$, that
$B_N\left (\Delta_m,f_J,
\left\lbrace a_i\right\rbrace ;T_c \right )$
is a sum of sinusoids at the same frequency, 
and that the sum of sinusoids of the same frequency
with possibly different phases and amplitudes is also a sinusoid at that frequency,
we have 
\be\label{eq_FN}
F_N \left (f_J, \left\lbrace a_i\right\rbrace ; T_c \right )
%
&=& 
\left|A_N \left (f_J;T_c \right )\right|
\nonumber \\
&& \quad \times \max_{\Delta_m}
\left|B_N\left
(\Delta_m,f_J,\left\lbrace
 a_i\right\rbrace ;T_c \right )\right| 
 \nonumber\\
 &=& 
\left|A_N \left (f_J;T_c \right )\right|
\sqrt{\mathcal{X}_N^2+\mathcal{Y}_N^2} 
\ee 
from \eqref{eq_R_N_STJ}-\eqref{eq_BN}, where
\be
F_N \left (f, \left\lbrace a_i\right\rbrace ; t \right )
= \max_{\alpha} \left| R_{N,T}
\left (\alpha; f, t \right )\right| ,
\ee
\be
\mathcal{X}_N=\sum_{i=1}^{N}{a_i \cos \frac{N-1+2i}{2N}\pi f_J T_c},
\ee
and
\be
\mathcal{Y}_N=\sum_{i=1}^{N}{a_i \sin \frac{N-1+2i}{2N}\pi f_J T_c} .
\label{eq-20}
\ee
%
Employing \eqref{eq_FN} in \eqref{eq_Opt_knwonF}, we can
express the problem of designing waveforms
with the estimated frequency $\widehat{f}_J$ as
\begin{equation}\label{eq_Opt_Sim_knwonF}
\begin{aligned}
\widehat{W}_{N,\widehat{f}_J}=&\arg 
\min_{\left\{ a_i \right\}_{i=1}^{N}}
{F_N \left ( \widehat{f}_J, \left\lbrace a_i\right\rbrace ; T_c \right )}\\
&\mbox{such that } \sum_{i=1}^{N} {a_i^2}=1,
\end{aligned}
\end{equation}
where
\be
\widehat{W}_{N,\widehat{f}_J}=
\left\{\hat{a}_k\right \}_{k=1}^{N}
 \ee
denotes a set of $N$ coefficients of $\widehat{w}_{N,\widehat{f}_J}(t)$,
an approximation of the optimal waveform
$\widehat{w}_{tr,\widehat{f}_J}(t)$
defined in \eqref{eq_Opt_knwonF} with $N$ rectangular pulses.

Let us in passing note that
\be
A_N \left (f;T_c \right )=0
\ee
if 
$f T_c=2kN$ for an integer $k$.
In addition, $A_N \left (f;T_c \right )$ decreases rather fast as $f$
increases beyond $ f= \frac{2N}{T_c} $. Therefore, we will concentrate on
the range 
from $0$ to $\frac{2N}{T_c}$ Hz of jamming frequency.


\subsection{Solutions and Algorithms}

Let us note that
the maximum $ F_N $ in \eqref{eq_Opt_Sim_knwonF} 
can now be rewritten as
\begin{equation}\label{eq:FN_expand}
\begin{aligned}
&F_N\left (\widehat{f}_J, \left\lbrace a_i\right\rbrace ; T_c\right)
=\left|A_N\left (\widehat{f}_J;T_c \right )\right|\\
&\times \sqrt{\sum_{i=1}^{N}{a_i^2}+2\sum_{n=1}^{N-1}
\cos{\frac{\pi n \widehat{f}_J T_c }{N}}{\sum_{k=1}^{N-n}{a_k a_{k+n}} }}\\
&=\left|A_N\left(\widehat{f}_J;T_c\right)\right|
\sqrt{ \mathbf{a}^T\,\mathcal{C} \mathbf{a}}
\end{aligned}
\end{equation}
from  \eqref{eq_FN}, where $  \mathbf{a}=\left[a_1, a_2, ..., a_{N}\right]^T $
and the matrix $\mathcal{C}$ has
%
\be
c_{i,j}=\cos \left(
\frac{\pi \widehat{f}_J T_c \left| i-j\right| }{N}\right)
\ee
as its  $(i,j)$-th elements.
The problem \eqref{eq_Opt_Sim_knwonF} of designing waveforms
with \eqref{eq:FN_expand} can thus be represented 
as a minimization of  $ \mathbf{a}^T\,\mathcal{C} \mathbf{a} $, for which
%
%
%
%
%
the solution should satisfy the Karush–-Kuhn–-Tucker (KKT)
conditions \cite{KKTcond}
\begin{equation}\label{eq:KKTcondition}
\mathbf{a}^T\left(\mathcal{C}-\lambda I \right) =0;
\qquad \mathbf{a}^T \mathbf{a}=1;
\qquad \lambda>0.
\end{equation}
%
%

The KKT conditions \eqref{eq:KKTcondition} imply that the solution
to the minimization of  $ \mathbf{a}^T\,\mathcal{C} \mathbf{a} $ will
be the normalized eigenvectors of the positive eigenvalues 
of  the matrix $ \mathcal{C} $.
The waveform solution to \eqref{eq_Opt_Sim_knwonF} for mitigating the STJ
can thus be obtained, for example, by
Powell's conjugate‐-direction method \cite{Powell}
after taking the conditions
\eqref{eq:KKTcondition} into account as
described in Algorithm 1, where $ \left[\mathbf{u}_1,
\mathbf{u}_2,...,\mathbf{u}_N \right] $ denotes a
set of initial direction vectors and $\mathbf{e}_i$
is the standard unit vector.

\begin{algorithm}[t]
\DontPrintSemicolon
\caption{Design of AJ waveforms against STJ with Powell's method}\label{alg:Powell}
\KwData{An estimate $ \widehat{f}_J$ of $ f_J $, $ h(\mathbf{a})
= \mathbf{a}^T\,\mathcal{C} \mathbf{a}$,\;
~~~~~~~~Initial $ \mathbf{a}_i $ }
\KwResult{$ \hat{\mathbf{a}} \leftarrow \mathbf{a}_i$}
\BlankLine
\While{$ \mathcal{C} \mathbf{a}_i \neq \lambda \mathbf{a}_i $ for some $
\lambda>0 $}{
$ \left[\mathbf{u}_1,\mathbf{u}_2,...,\mathbf{u}_N \right] \leftarrow
\left[\mathbf{e}_1,\mathbf{e}_2,...,\mathbf{e}_N \right] $.\;
$i \leftarrow 1$.\;
Initialize $ \mathbf{a}_i $.\;
\While{not converged}{
$ \mathbf{p}_0 \leftarrow \mathbf{a}_i $\;
\For{$ k=1,2...N $}{$ \hat{\gamma}_k \leftarrow \arg\min_{\gamma_{k}}
{h\left (\mathbf{p}_{k-1}+\gamma_k \mathbf{u}_k \right)} $.\;
$ \mathbf{p}_k \leftarrow \mathbf{p}_{k-1} +\hat{\gamma}_k \mathbf{u}_k$. }
$ i \leftarrow i+1 $.\;
\For{$ j=1,2,...,N-1 $}{$ \mathbf{u}_j \leftarrow \mathbf{u}_{j+1} $.}
$ \mathbf{u}_N \leftarrow \mathbf{p}_N- \mathbf{p}_0 $.\;
$ \hat{\gamma} \leftarrow \arg\min_{\gamma}{h\left (\mathbf{p}_{0}+\gamma
\mathbf{u}_N \right )} $.\;
$ \mathbf{a}_i \leftarrow \mathbf{p}_0 + \hat{\gamma} \mathbf{u}_N $.
}
}
\end{algorithm}

\subsection{Challenges and Possible Solutions in Other Jamming Scenarios}
The proposed design of waveforms provides us with an optimality against STJ; yet,
we will probably encounter in practice with other more complex and intelligent jamming scenarios.
Let us briefly describe how we can generalize and extend the proposed waveform design
for coping with
the challenges in such various jamming scenarios.

Many intelligent jamming (IJ), such as the MTJ,
TV-STJ, and SWJ as mentioned before, schemes are based on
the concepts of optimization, awareness, game theory, and software defined communications
for the efficiency and effectiveness of jammers.
With the knowledge of the protocol, IJ schemes are shown to
perform (from the viewpoint of jammers) 
better than the trivial continuous high power noise
jamming while also retaining its effectiveness:
More specifically, IJ schemes are reported to be more efficient
than the periodic and trivial jammings by one to two and five, respectively,
orders of magnitude  \cite{CounterIJ}.
%

Now, although
application of the proposed waveforms against STJ directly to
the MTJ cases may not be promising,
the design of waveforms against MTJ can be formulated, for example, as
\begin{equation}\label{eq_Opt_knwonMF}
\begin{aligned}
\widehat{w}_{tr,\widehat{\mathcal{F}}_J}(t)=&\arg \min_{w_{tr}}
\max_{\Delta_m} \sum_{\omega=1}^{\Omega}\left|R_{tr,T}
\left (\Delta_m; \widehat{f}_{J,\omega}, T_c \right )\right|\\
&\mbox{such that }\left\| \widehat{w}_{tr}(t)\right\|^2=1
\end{aligned}
\end{equation}
by extending \eqref{eq_Opt_knwonF} to take the multiple estimates
\be
\widehat{\mathcal{F}}_J=\left\lbrace \widehat{f}_{J,\omega}
 \right\rbrace_{\omega=1}^{\Omega}
\ee
of jamming frequencies into consideration.
We expect that simplification and/or approximation of the problem \eqref{eq_Opt_knwonMF}
for the design of waveforms against MTJ
would provide us with some insight or solution for possible practical implementation.
We would also like to note that
MTJ with the jamming frequencies falling inside (a small portion of) the bandwidth
of the TH SS systems can statistically be considered as a STJ in the perspective of the AJ performance \cite{NBI-MPC-Anal}.

The TV-STJ is a subclass of random jamming, in which 
several tones varying in time are employed \cite{FH-PAC}.
The waveforms designed by the proposed technique can clearly be
employed within periods the TV-STJ. 
When the variation of the TV-STJ
is very fast, the problem could possibly be
modeled and solved with
%
the methods of signal detection and classification,
%
widely developed by utilizing energy detectors, higher-order statistical features,
and statistical tests \cite{SigDetClass}: They can also be achieved by
using learning methods based on 
techniques of neural networks
as shown in 
\cite{JamClassCNN, SmartJamClass}.

For SWJ, which swepdf a narrow-band jamming signal over a wide frequency band,
sinusoidal signals are widely used with a linear sweep method
because of
simplicity and usefulness \cite{SWJ}.
By viewing the SWJ as a TV-STJ, the technique described above against the TV-STJ
can similarly be applied against SWJ by optimizing waveforms periodically. Alternatively,
by generalizing 
\eqref{eq_Opt_knwonF} again,
we can design a waveform against the SWJ over the bandwidth of the SWJ as
\begin{equation}\label{eq_Opt_knwonSWF}
\begin{aligned}
\widehat{w}_{tr,\widehat{\mathcal{F}}_J}
=&\arg \min_{w_{tr}} \max_{\Delta_m} \int_{\widehat{f}_{J,l}}^{\widehat{f}_{J,u}}\left|R_{tr,T}
\left (\Delta_m; f, T_c \right )\right| df\\
&\mbox{such that }\left\| \widehat{w}_{tr}(t)\right\|^2=1,
\end{aligned}
\end{equation}
for example, where $ \widehat{\mathcal{F}}_J=\left\lbrace \widehat{f}_{J,l} , \widehat{f}_{J,u}
 \right\rbrace$ now
denotes
the set of estimates
of the lower and upper jamming frequencies 
of the SWJ.
The problem \eqref{eq_Opt_knwonSWF} is applicable to the partial-band noise jamming
signal as well, but is rather intractable for a direct solution
mainly due to the minimization, maximization,
and integration over continuous spaces involved in the optimization.
Apparently, simplification and/or approximation of the problem \eqref{eq_Opt_knwonSWF}
would constitute a good topic for a future study.



\section{Numerical and Simulation Results}
\label{sec-simul}

%
%
In this section, we discuss the performance of the proposed method of designing waveforms
via numerical and simulation results for $N=5$ and $T_c = 1$ ns with
the pulse duration $ T_p=T_c/2=0.5 $ ns and approximate bandwidth of $5$ GHz as in
\cite{THseq-Design}. 
Normally, TH SS systems are deployed in network applications; yet, to focused fully on the
effects of jammer \cite{NBI-MPC-Anal}, 
we consider scenarios with one-node-only and static setup without taking
mobility, Doppler's effect, and inter-node interference into consideration.


\subsection{Numerical Results}

From \eqref{eq_AN}-\eqref{eq-20}, 
we have
\begin{equation}\label{eq_CostFn_5ps}
F_5(f_J, \left\lbrace a_i\right\rbrace ; 1 
)=\left|A_5(f_J;1 
)\right|
 \sqrt{\mathcal{X}_5^2+\mathcal{Y}_5^2} ,
\end{equation}
where
\begin{equation}\label{eq:A5}
A_5(f;1 
)= \frac{2}{\pi f} \sin\frac{\pi f}{2} \sin\frac{\pi f}{10},
\end{equation}
\begin{equation}\label{eq:X5}
\mathcal{X}_5=\sum_{i=1}^{5}{a_i \cos \frac{i+2}{5}\pi f_J},
\end{equation}
\begin{equation}\label{eq:Y5}
\mathcal{Y}_5=\sum_{i=1}^{5}{a_i \sin \frac{i+2}{5}\pi f_J},
\end{equation}
and $ f_J $ is in GHz.


\begin{figure}[t!]
	\centering
	\includegraphics[width = 8.5cm]{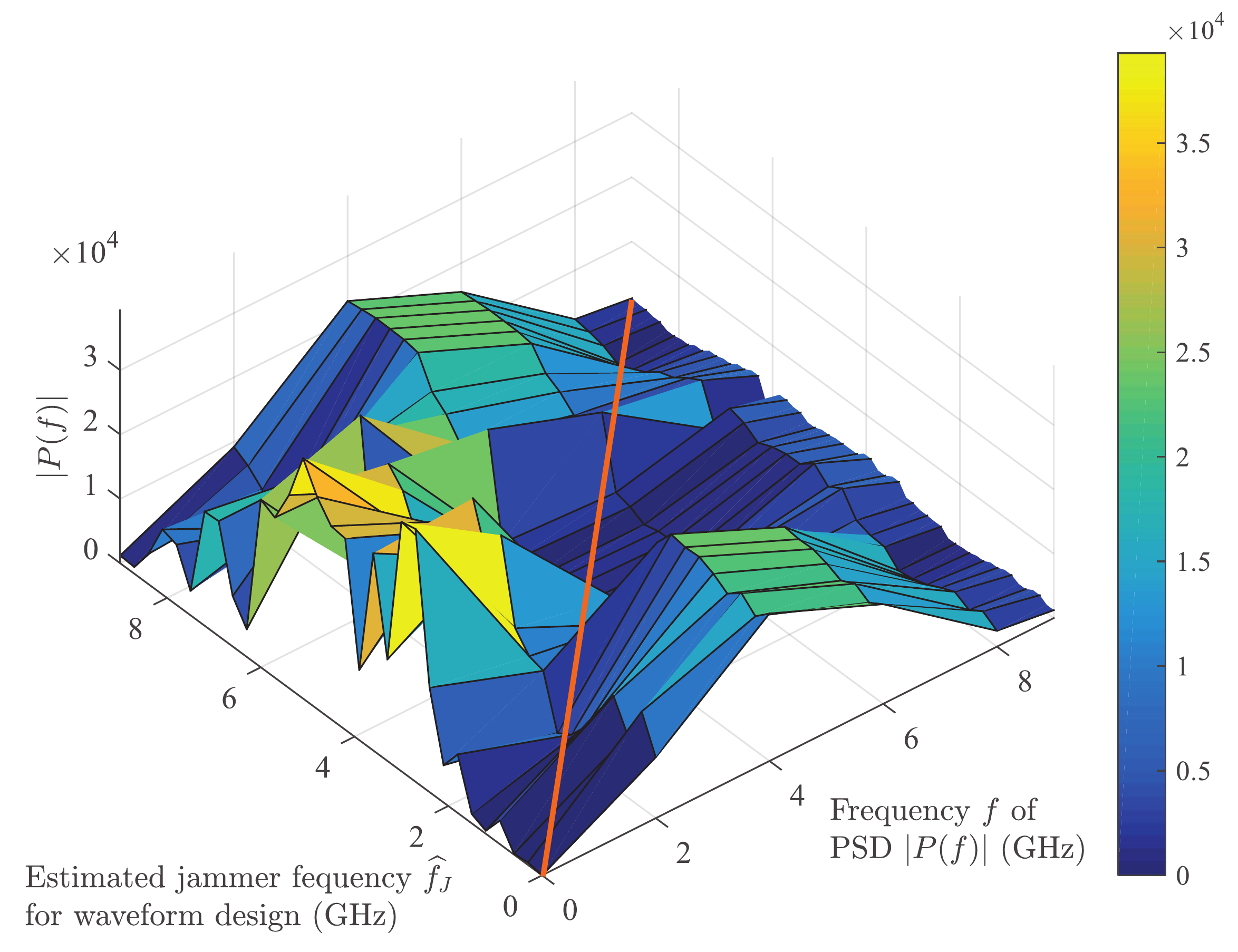} 
	\caption{Spectrogram of the optimal waveforms. Each row represents the
PSD of the waveform optimized
for the estimated jammer frequency $\widehat{f}_J$.}\label{fig:Opt_PSD}
\end{figure}

Fig. \ref{fig:Opt_PSD} shows the spectrogram of the waveform optimized for
the estimated jammer frequency $\widehat{f}_J$, where
the color density 
represents the PSD of the waveform.
It is clearly observed that the lowest 
PSD is located at the frequency for which the waveform is optimized: i.e.,
colors on the diagonal line are generally darker blue (meaning lower PSD
values) than those away from the diagonal line. 

Let us next consider the BER performance of some waveforms with
simulation parameters 
$N_f=3$, $N_c=4$, $N_p=100$, $T_f=4$ ns, $T_m=0.25$ ns,
$T_p=0.5$ ns, $\delta=0.5$ ns, and a
sampling interval of 
0.02 ns for all signals \cite{NBI-MPC-Anal}: We have chosen $T_f = 4$ for simplicity in the simulations
although it does not 
completely satisfy the condition $T_f \gg T_c$ of practical cases.
\begin{figure}[t!]
	\centering
	\includegraphics[width = 8.5cm]{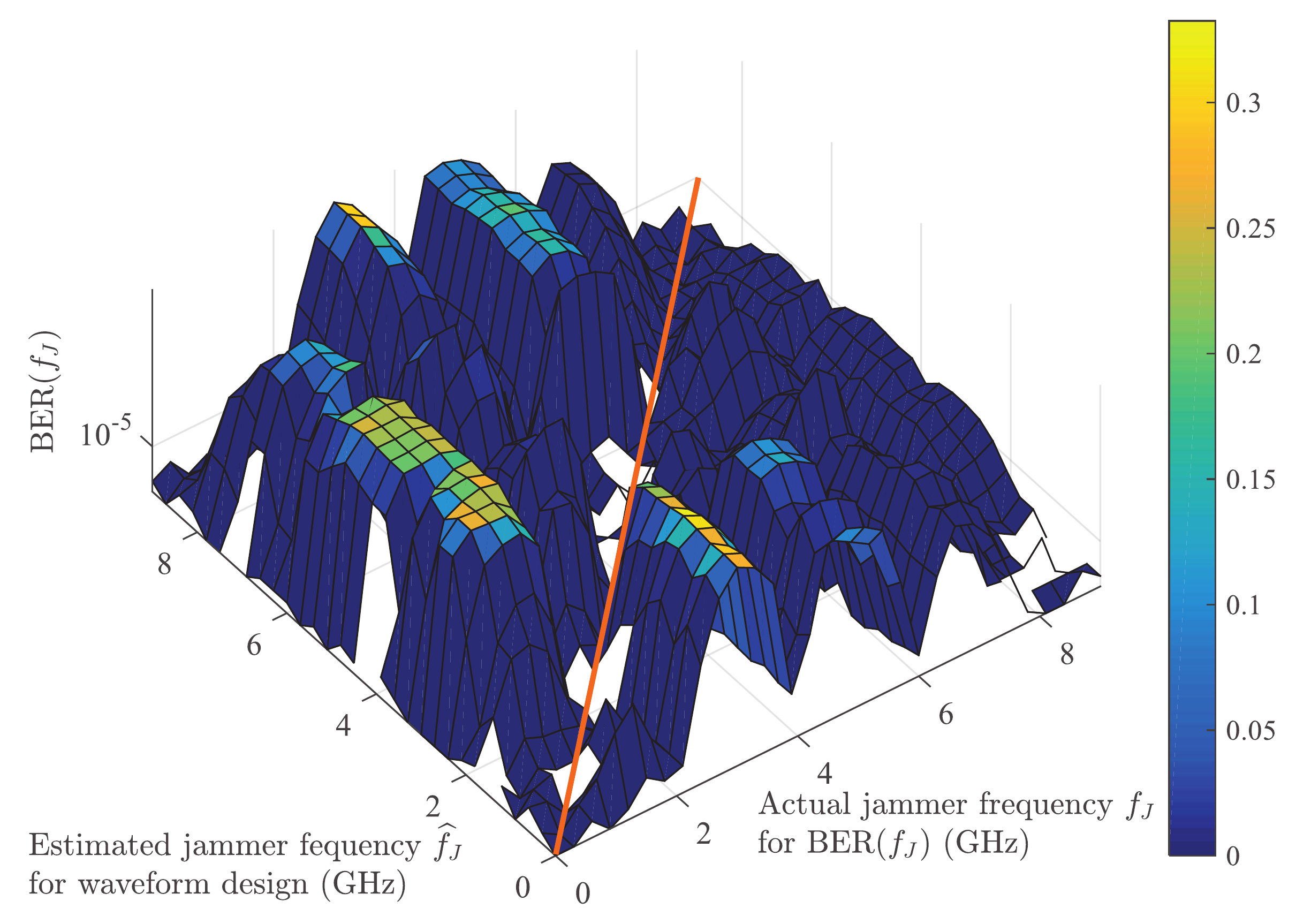} 
	\caption{BER performance of the optimal waveform at $f_J$ versus
$f$.}\label{fig:Opt_BER_spec}
\end{figure}
Fig. \ref{fig:Opt_BER_spec} simulates the BER performance of the TH system
with the waveforms optimized for the estimated jammer frequency $\widehat{f}_J$
versus the actual jammer frequency $f_J$. The BER performance along the diagonal
line is very close to the minimum simulated BER of
$2.5 \times 10^{-7}$: This implies
that the proposed design of waveforms improves the BER performance
of the TH system when the estimate $\widehat{f}_J$ is close to $f_J$.
%
%
As the BER is observed to decrease sharply
over 8 GHz, we will mainly focus on the interval $[0, 9]$ GHz
from now on. 

\begin{figure}[t!]
	\centering
	\includegraphics[width = 8.5cm]{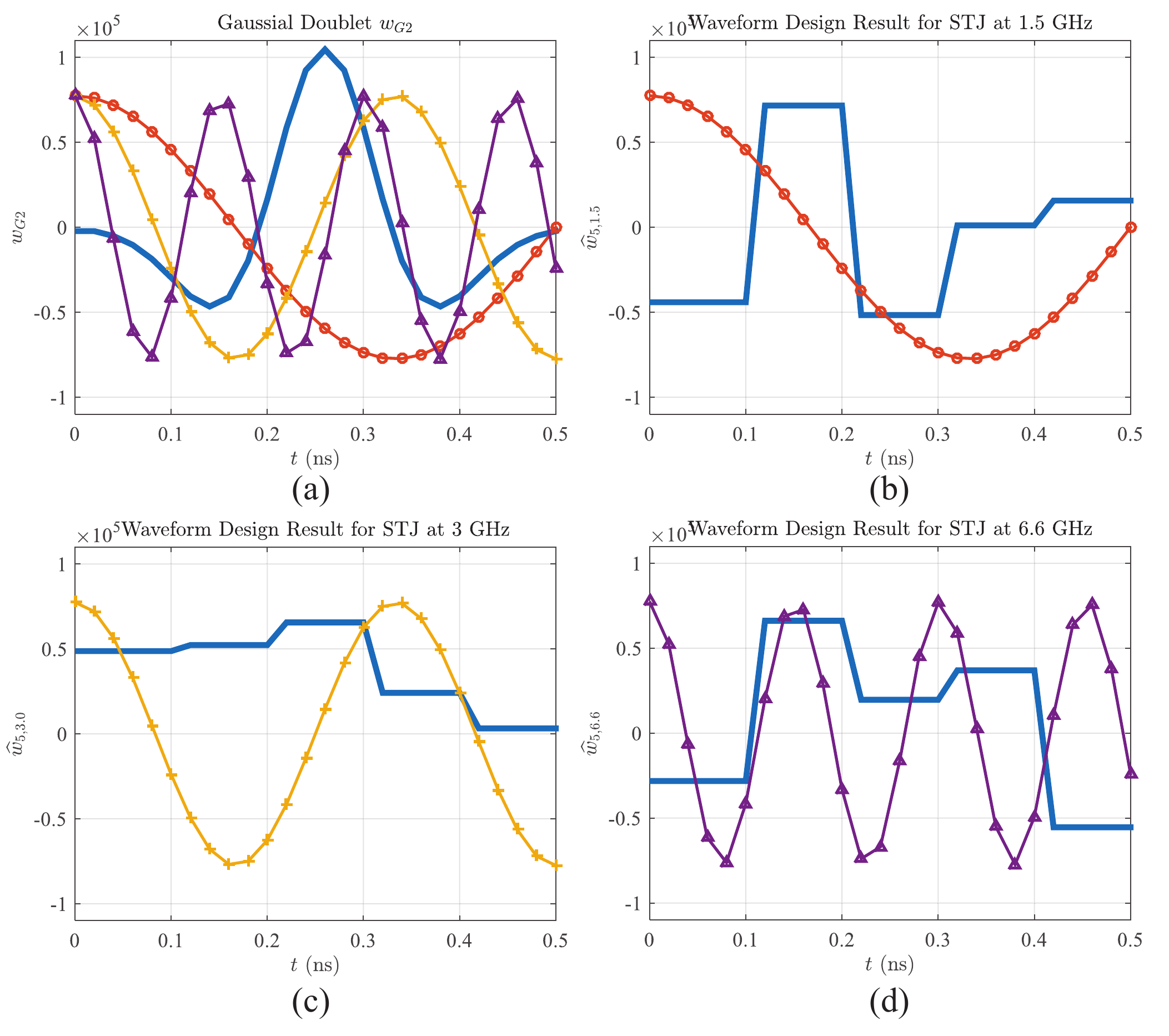} 
	\caption{Proposed waveforms of the TH system (the blue curve) with
$ N=5 $ against STJ signals of SJR=$0$ dB at 1.5 (the red curve with
`O' markers), 3.0 (the yellow curve with `+' markers), and 6.6
GHz (the purple curve with `$\triangle$' markers): (a)
$ w_{G2} $, (b) $\widehat{w}_{5,\textrm{1.5}}$, (c)
$\widehat{w}_{5,\textrm{3.0}}$, and (d)
$\widehat{w}_{5,\textrm{6.6}}$.}\label{fig:Opt_pulses}
\end{figure}

In the following simulations, we mainly show
the results for
the actual jamming frequency $f_J$
of $1.5$, $3.0$, and $6.6$ GHz for a clarity reason
in the figures: Let us mention that we have nonetheless
performed simulations from 0 to 9 GHz in an interval of 0.3 GHz.
The AJ performance of the TH systems with the optimized waveforms
$\widehat{w}_{5,1.5}(t)$, $\widehat{w}_{5,3.0}(t)$, and $\widehat{w}_{5,6.6}(t)$
are compared with that of the TH system with the conventional waveform and
that of the
TH system with the clipper receiver described by
the threshold \eqref{eq:clipper}.
Here, the `conventional' waveform we considered 
is the Gaussian doublet
\be
w_{G2}(t+0.25)= A\left( 1-64\pi t^2 \right) \exp\left(-32\pi t^2 \right)
\ee
with the unit of $t$ in ns \cite{THUWB-00}, and
the optimized waveforms $\widehat{w}_{5,1.5}(t)$,
$\widehat{w}_{5,3.0}(t)$, and $\widehat{w}_{5,6.6}(t)$ can be expressed
by 
\begin{equation}
\widehat{W}_{5,\textrm{1.5}}=\left\lbrace -0.441,\, 0.717,\, -0.517, 0.013,\,
0.157 \right\rbrace \label{eq_w5_15},
\end{equation}
\begin{equation}
\widehat{W}_{5,\textrm{3.0}}=\left\lbrace 0.487,\, 0.523,\, 0.656,\, 0.241,\,
0.032 \right\rbrace \label{eq_w5_30},
\end{equation}
and
\begin{equation}
\widehat{W}_{5,\textrm{6.6}}=\left\lbrace -0.282,\, 0.662,\, 0.197,\, 0.370,\,
-0.554 \right\rbrace , \label{eq_w5_66}
\end{equation}
respectively, as obtained
by solving \eqref{eq_Opt_Sim_knwonF} numerically with Algorithm \ref{alg:Powell}.
The waveforms $ w_{G2}(t) $, $\widehat{w}_{5,\textrm{1.5}}(t)$,
$\widehat{w}_{5,\textrm{3.0}}(t)$, and $\widehat{w}_{5,\textrm{6.6}}(t)$
are shown in Fig. \ref{fig:Opt_pulses} with STJ signals at 1.5, 3.0, and 6.6 GHz.

\begin{figure}[t!]
	\centering
	\includegraphics[width = 8.5cm]{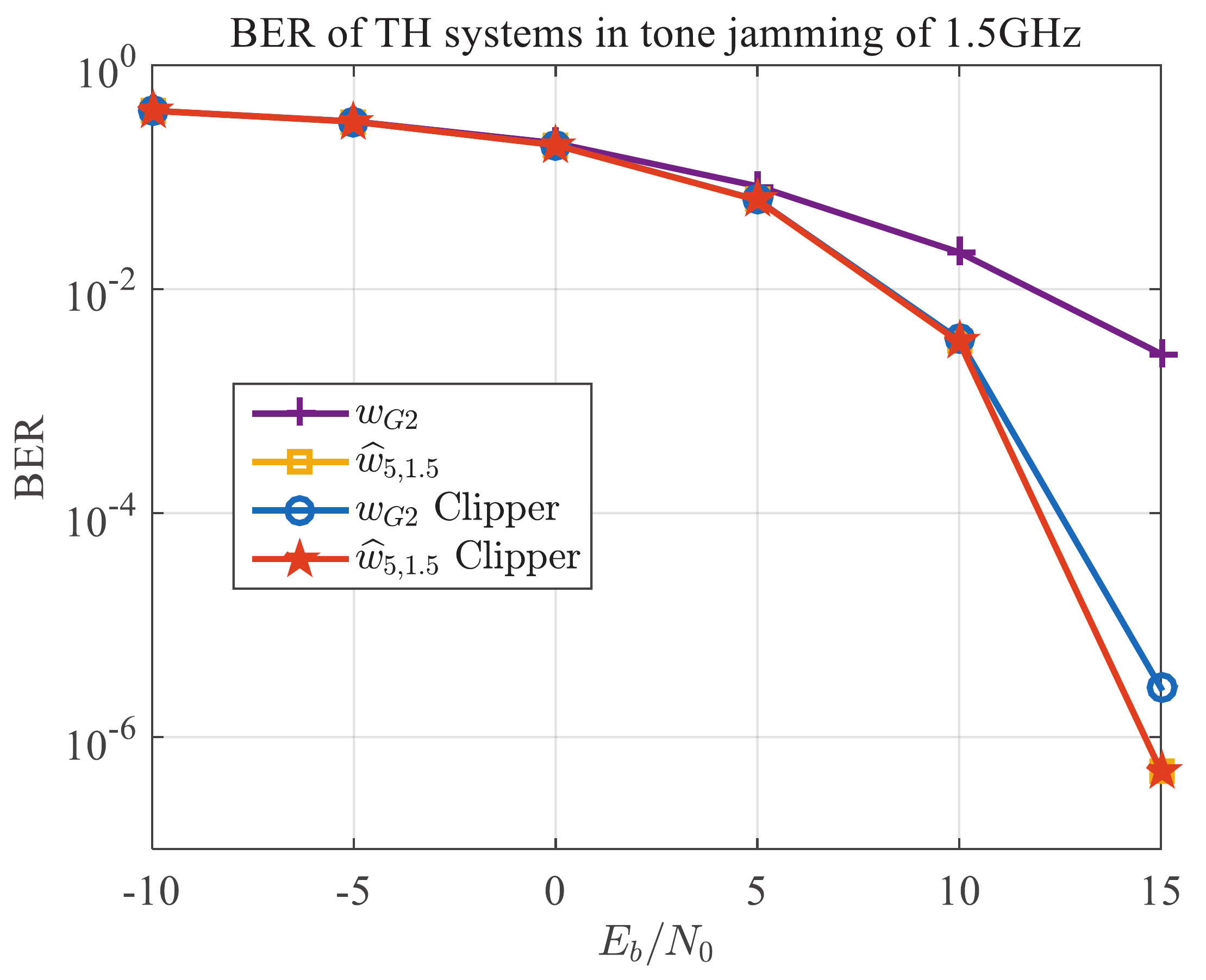} 
	\caption{Comparison of BER of the TH systems
with $\widehat{w}_{5,\textrm{1.5}}$ (the yellow curve with `$\Box$' markers),
with $\widehat{w}_{5,\textrm{1.5}}$ and the clipper
(the red curve with `$\medstar$' markers),
with $ w_{G2} $ (the purple curve with `+' markers), and
with $ w_{G2} $ and the clipper (the blue curve with `O' markers)
in an AWGN channel with STJ of SJR=$-10$ dB at 1.5 GHz.}  \label{fig:BER_15GH}
\end{figure}

In Fig. \ref{fig:BER_15GH}, we compare the BER of the TH systems employing
$\widehat{w}_{5,\textrm{1.5}}$ (the yellow curve with `$\Box$' markers),
$\widehat{w}_{5,\textrm{1.5}}$ with clipping (the red curve with
`$\medstar$' markers),
$ w_{G2} $ (the purple curve with `+' markers),
and
$ w_{G2} $ with clipping (the blue curve with `O' markers)
over an AWGN channel with
signal-to-jamming ratio (SJR) of $-10$ dB and $ f_J=1.5 $ GHz.
The BER of the TH system with the optimized waveform $\widehat{w}_{5,\textrm{1.5}}$
is clearly lower than that with the conventional Gaussian doublet $ w_{G2} $
even with clipping.

Additionally, $\widehat{w}_{5,\textrm{1.5}}$ exhibits
the same BER performance regardless of the clipper at the receiver.
This implies that the clipper does not provide 
an additional improvement of
performance beyond the optimization of waveform design.
We have confirmed that the same observation
holds when the jammer frequency is of other values
between 0 and 9 GHz although
not specifically shown in this figure for brevity.
In order to investigate the advantages of the proposed
design technique of waveforms conservatively, we assume the clipper receiver is
embedded in TH systems from now on.

%

\subsection{Simulation Results}

\begin{figure}[t!]
	\centering
	\includegraphics[width = 8.5cm]{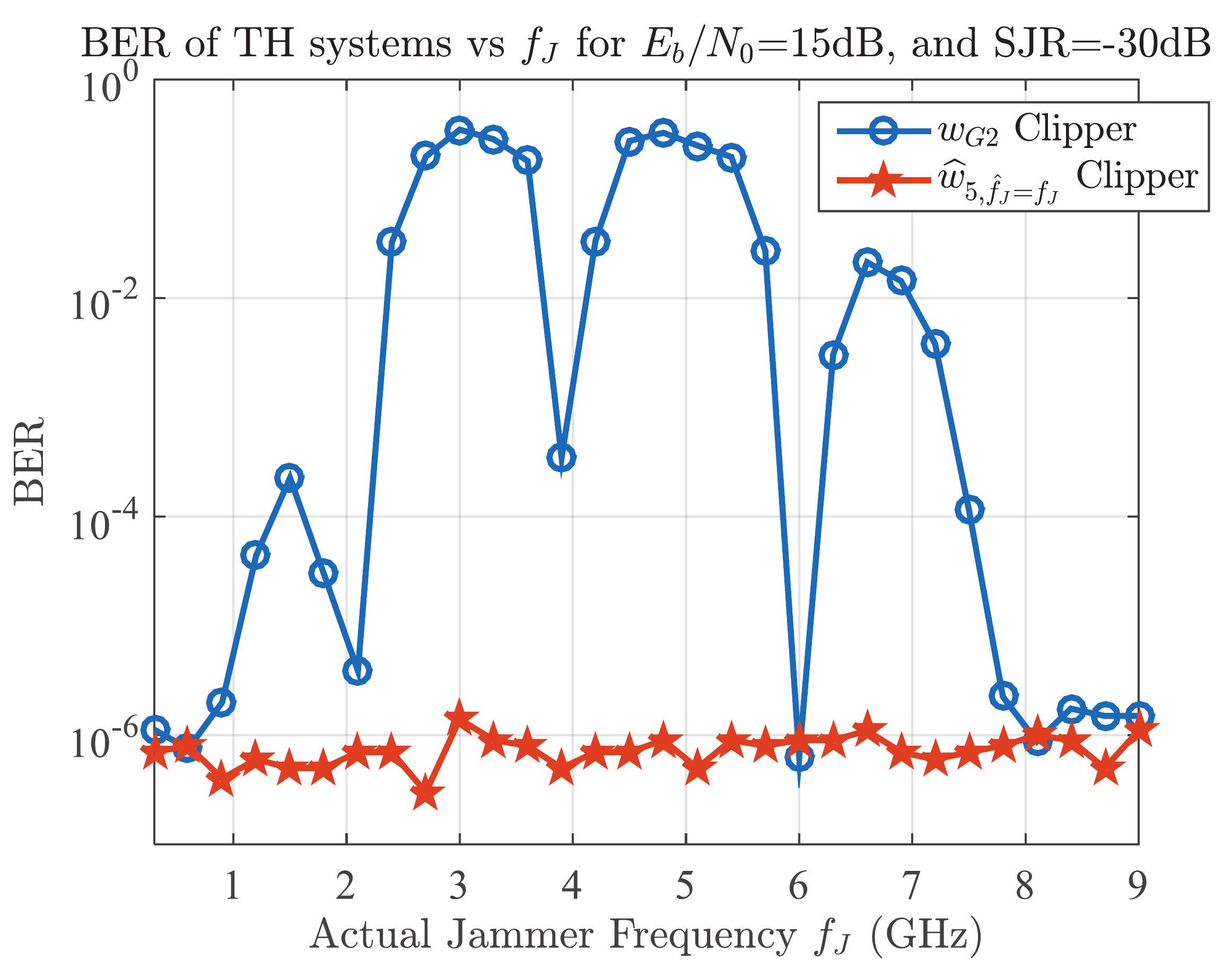} 
	\caption{BER performance versus jammer frequency $ f_J $ of
the TH systems employing the optimized waveform
$\widehat{w}_{5,\widehat{f}_J=f_J}$ (the red curve with
`$\medstar$' markers) and Gaussian doublet $ w_{G2} $
(the blue curve with `O' markers) over an AWGN channel of $ E_b/N_0 =15$ dB and
SJR=$-30$ dB.}\label{fig:BER_fJ_m30dB}
\end{figure}

Fig. \ref{fig:BER_fJ_m30dB} presents the BER performance versus the actual
jammer frequency $ f_J $ of the TH systems employing
the optimized waveform
$\widehat{w}_{5,\widehat{f}_J=f_J}$ (the red curve with `$\medstar$' markers)
and Gaussian doublet $ w_{G2} $ (the blue curve with `O' markers)
over an AWGN channel with the
bit energy to noise power spectral density ($ E_b/N_0 $) $15$ dB and SJR=$-30$ dB.
At almost all values from 0 to 9 GHz of the jammer frequency,
the simulation results again
show that the optimized waveforms generally outperform the
conventional scheme with clipping.
The improvement of AJ performance with the optimized waveforms
over the Gaussian doublet 
becomes very large when the jammer frequency is 3, 5, and 6.5 GHz,
at which the Gaussian doublet has most of its power.

\begin{figure}[t!]
	\centering
	\includegraphics[width = 8.5cm]{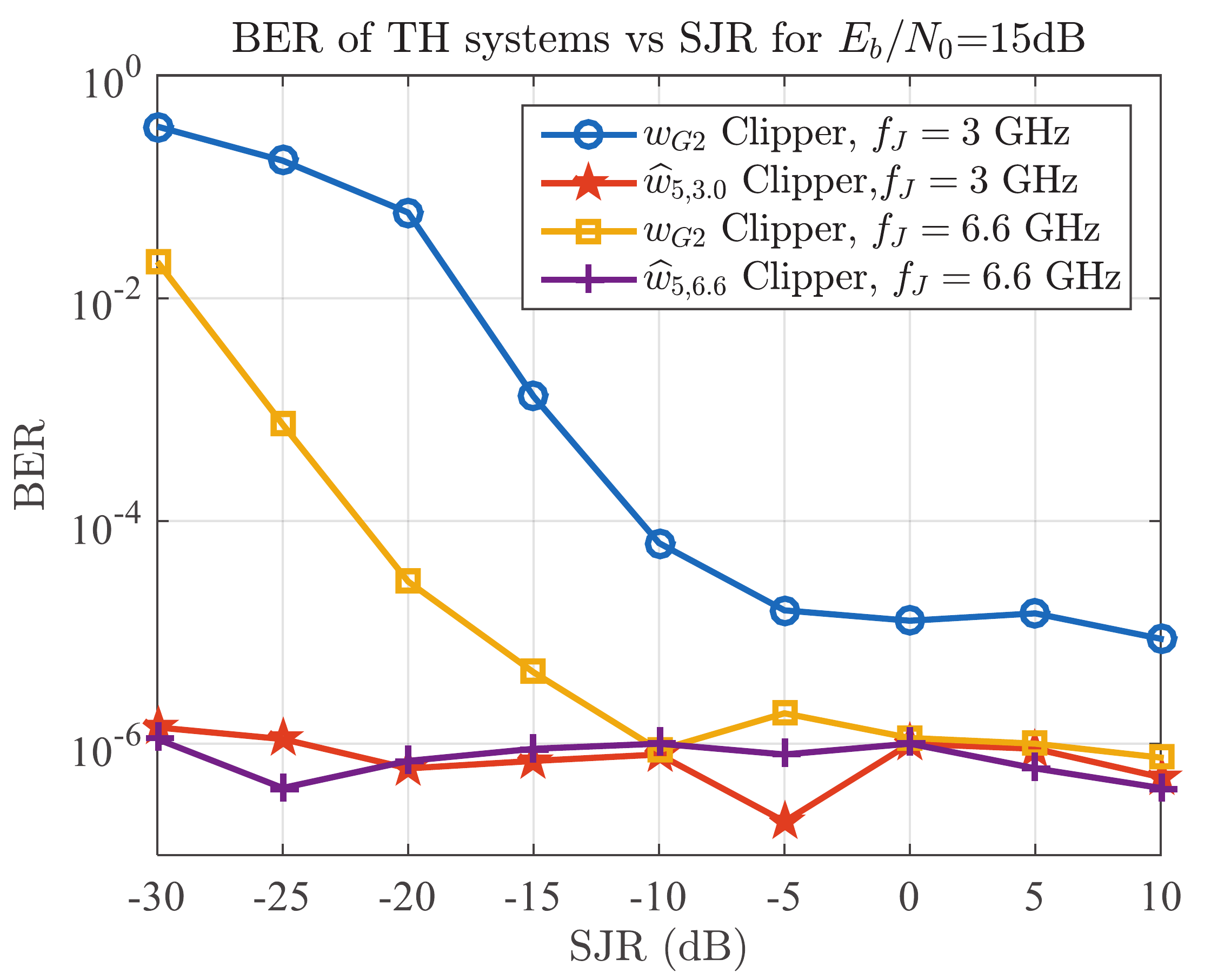} 
	\caption{BER performance of the TH systems with
the optimized waveform $\widehat{w}_{5,3.0}$
(the red curve with `$\medstar$' markers) and
Gaussian doublet $ w_{G2} $ (the blue curve with `O' markers)
against an STJ at 3 GHz, and
the optimized waveform $\widehat{w}_{5,6.6}$ (the purple curve with
`+' markers) and
Gaussian doublet $ w_{G2} $ (the yellow curve with
`$\Box$' markers)
against an STJ at 6.6 GHz.}
\label{fig:BER_SJR_30GH}
\end{figure}

In Fig. \ref{fig:BER_SJR_30GH}, we consider the BER performance
of the TH systems
with the optimized waveform $\widehat{w}_{5,3.0}$
(the red curve with `$\medstar$' markers)
and
with the Gaussian doublet $ w_{G2} $ (the blue curve with `O' markers)
 for the STJ at $ f_J= 3$ GHz
when $ E_b/N_0 =15$ dB. The BER of the TH systems
with $ w_{G2} $ (the yellow curve with `$\Box$' markers) and
with $\widehat{w}_{5,6.6}$ (the purple curve with `+' markers)
for the STJ at $ f_J=6.6$ GHz are also shown.
We observe that the optimized waveforms $ \widehat{w}_{5,3.0} $
and $ \widehat{w}_{5,6.6} $ provide a stable AJ performance of BER=$ 10^{-6}$
even when the SJR 
of the STJ varies.
On the other hand, the BER of the TH system with $ w_{G2} $ becomes
worse as the SJR decreases. In addition, when the SJR increases,
the BER performance of $ w_{G2}$ for $f_J = 3.3 $ GHz is saturated
at the level of $10^{-5}$, a value (roughly 10 times) higher
than that of the proposed waveform.
We have confirmed that the same observation
holds at other values in the range $[0, 9]$ GHz of 
the jammer frequency. 

\subsection{Imperfect Estimation of Jamming Frequency}

We have so far assumed
idealistically perfect estimation of $ f_J $, which is not always possible
in
practical scenarios especially when the jammer power is not strong enough.
Let us now consider the scenario that an estimation error
\be
\varepsilon_f=\widehat{f}_J -f_J
\ee
occurs in estimating the jammer frequency $ f_J $.
%
The estimation error $ \varepsilon_f $ is commonly assumed to follow 
a Gaussian distribution, i.e.,
%
\be
\varepsilon_f \sim \mathcal{N}\left (\mu_{\varepsilon_f},
\sigma_{\varepsilon_f}^2 \right ) ,
\ee
where $ \mu_{\varepsilon_f} $
and $ \sigma_{\varepsilon_f} $ are the mean and standard deviation of
$ \varepsilon_f $, respectively \cite{CFO}.

\begin{figure}[t!]
	\centering
	\includegraphics[width = 8.5cm]{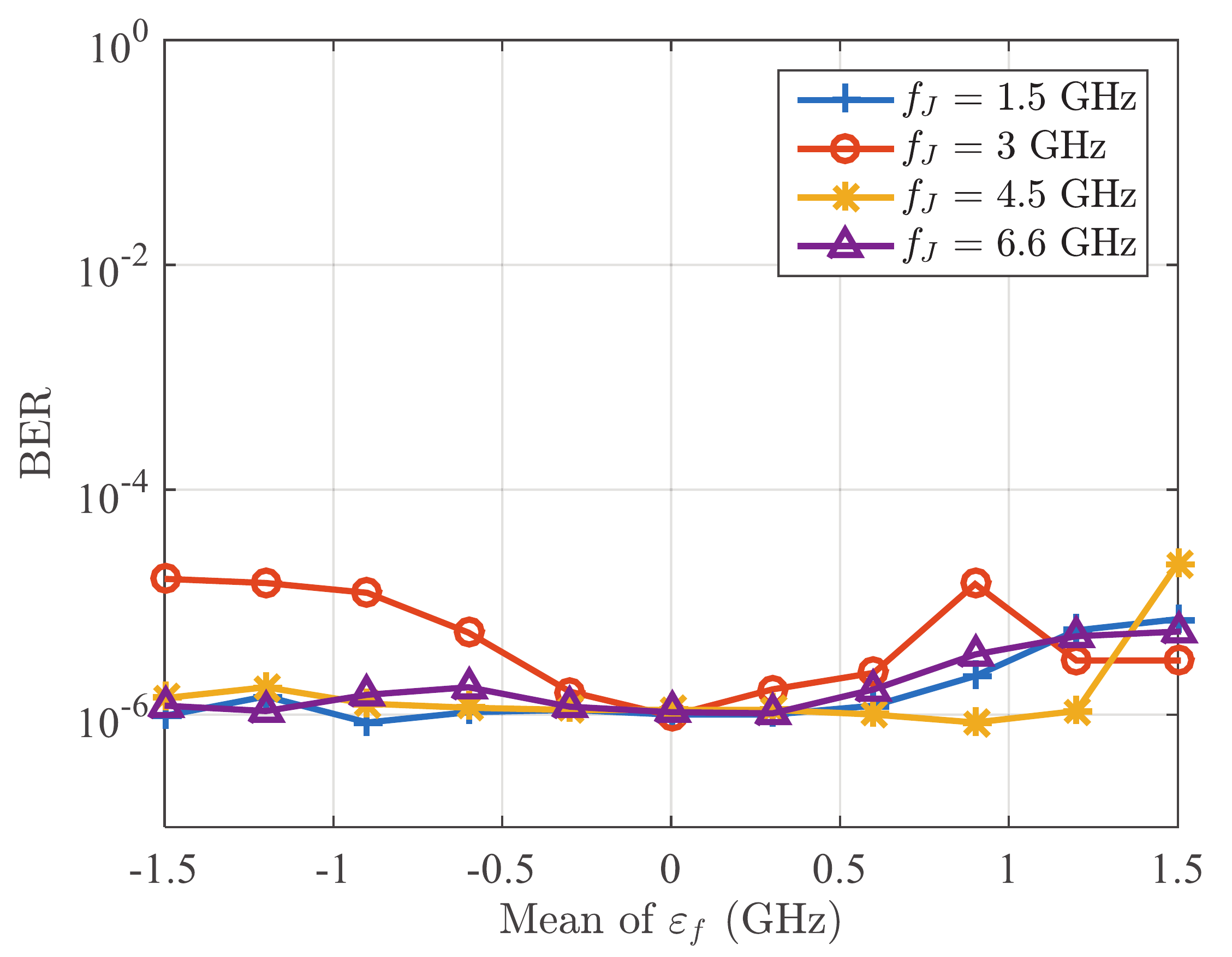}
	\caption{BER performance of the TH systems employing
$\widehat{w}_{5,\widehat{f}_J}$ versus the mean
of the estimation error $ \varepsilon_f = \widehat{f}_J-f_J$ when
the standard deviation of the estimation error is 0
over an AWGN
channel of $ E_b/N_0=15 $ dB and STJ of
SJR=$-10$ dB at 1.5 (the blue curve with `+' marker), 3 (the red curve with
`O' markers), 4.5 (the yellow curve with `*' markers), and 6.6 GHz (the purple curve with `$\bigtriangledown$' markers).}\label{fig:BER_MEAN_df}
\end{figure}

Fig. \ref{fig:BER_MEAN_df} shows the BER performance of the TH
system with $\widehat{w}_{5,\widehat{f}_J}$ versus
$\mu_{\varepsilon_f}$ when 
$\sigma_{\varepsilon_f}=0$ for several values of STJ frequency.
The simulation results in this figure
indicate that the proposed design of waveforms provides
a reasonable BER level of
$2 \times 10^{-5}$ even when the mean of the estimation error is
from -1.5 to 1.5 GHz.
When $ \mu_{\varepsilon_f}=-1.5$ and $ 0.9 $ GHz, the AJ performance
of the TH system
with the proposed design of waveforms degrades mostly under the STJ with
$ f_J=3 $ GHz: Yet, 
even this most severe degradation
provides a BER of $10^{-5}$ approximately,
a value much better than the BER $8 \times 10^{-5}$
with $ w_{G2} $ (shown in Fig. \ref{fig:BER_SJR_30GH}).
In the cases of 1.5 and 6.6 GHz of $ f_J $, the estimation error from -1.5 to
0.3 GHz does not influence the AJ performance of the TH systems
with the proposed design of waveforms significantly.
The proposed design of waveforms
also exhibits a robustness property against estimation errors from
$-1.5$ to $1.2$ GHz for an STJ of $f_J=4.5 $ GHz.

\begin{figure}[t!]
	\centering
	\includegraphics[width = 8.5cm]{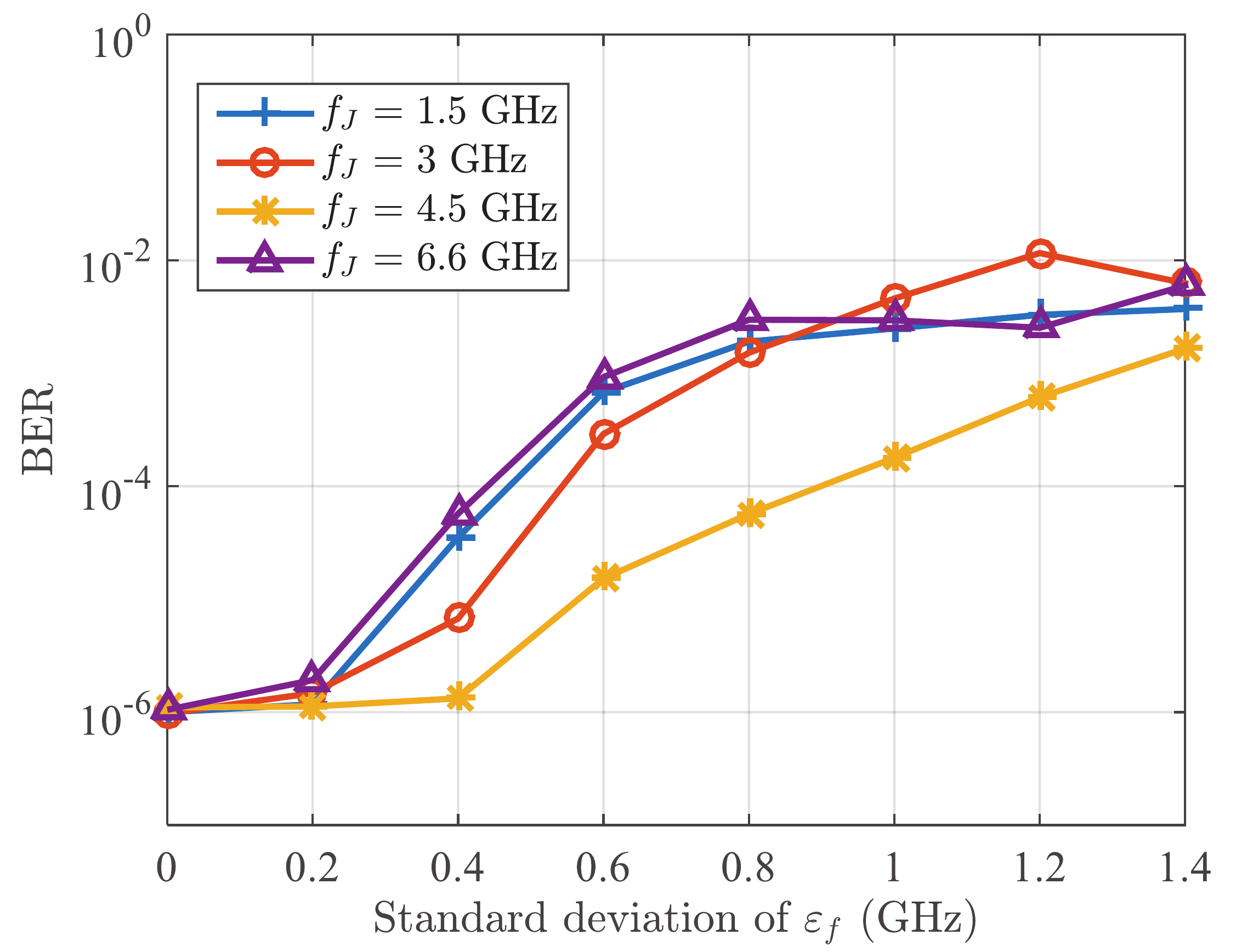}
	\caption{BER performances of the TH systems employing
$\widehat{w}_{5,\widehat{f}_J}$
versus the standard deviation 
of the estimation error
$ \varepsilon_f = \widehat{f}_J-f_J$ over an AWGN channel of $ E_b/N_0=15 $
dB and STJ of SJR=$-10$ dB at 1.5 (the blue curve with `+' marker),
3 (the red curve with `O' markers), 4.5 (the yellow curve with `*' markers),
and 6.6 GHz (the purple curve with `$\bigtriangledown$'
markers).}\label{fig:BER_STD_df}
\end{figure}

Fig. \ref{fig:BER_STD_df} presents the BER performance of the TH system with
$\widehat{w}_{5,\widehat{f}_J}$ versus $ \sigma_{\varepsilon_f} $ when 
$ \mu_{\varepsilon_f}=0  $ under STJ with various values of jamming frequency.
It is observed that
$ \sigma_{\varepsilon_f} $ of the estimator should be less than  0.46, 0.54, 0.9,
and 0.43 GHz when $ f_J=1.5, 3, 4.5, $ and $ 6.6  $ GHz, respectively,
to ensure a BER level of $10^{-4}$.
The proposed design of waveforms clearly 
provides a BER performance of $10^{-5}$
when the standard deviation of the estimation error is less than 0.3 GHz
for almost all values from 0 to 9 GHz of jammer frequency
although we have 
shown 
the results
only for $ f_J=1.5, 3, 4.5, $ and $ 6.6  $ GHz
for a brevity reason in this figure. 
In addition, we have considered the influence
of the mean and variance of the estimation error
only for limited cases: Yet, we believe
%
that the standard deviation $ \sigma_{\varepsilon_f} $ of the estimation error
is more crucial 
than the mean $ \mu_{\varepsilon_f} $ of the estimation error
when estimating the jammer frequency $ f_J $ in the design of AJ waveforms.

\section{Conclusion}
\label{sec-concl}
In this paper, the problem of designing waveforms with an aim of improving
the anti-jamming performance
against 
STJ has been addressed.
The problem of designing waveform is formulated and simplified 
by analyzing the correlation between
the TH and jamming signals.
Assuming an estimate of the frequency of STJ signal is available,
an algorithm is provided for the design 
of suboptimal waveforms. 

The 
waveforms designed with a perfect estimate
of the jamming frequency
outperform the conventional Gaussian doublet regardless
of the clipper for almost all values from 0 to 9 GHz of jammer frequency.
%
%
%
%
We have in addition observed that
the proposed design can provide us with waveforms
that overcome the unavoidable saturation of the
BER performance of the conventional Gaussian waveform even with a clipper.

In the case of non-ideal estimation of the jamming frequency,
simulation results showed that the AJ capability
the waveforms designed with the proposed scheme
still maintain 
a reasonable BER level of $2 \times
10^{-5}$ even when the mean of estimation error is in the range of $[-1.5 , 1.5]$
GHz.
In addition, the proposed design of waveforms can provide
a BER performance of $10^{-5}$
when the standard deviation of the estimate is less than 0.3 GHz for almost
all values from 0 to 9 GHz of jammer frequency.
We have also observed that the standard deviation
of the estimation error from the estimation of jammer frequency
is more influential 
than the mean 
of the estimation error 
in the design of AJ waveforms.
%
Finally, the proposed design of waveforms exhibits
robustness to the estimation errors 
of the jammer frequency.

We wish to add that consideration of complex and intelligent jamming scenarios (including
the MTJ, TV-STJ, and SWJ) in the design of waveforms, and investigation of joint
optimization of power allocation schemes and waveform design
are expected to be highly promising and fruitful topics for further studies.


\section*{Acknowledgments}
The authors would
like to thank the Associate Editor and three anonymous reviewers
for their constructive suggestions and helpful comments.

\input{ref-jhy-p1-is-21}


\begin{IEEEbiography}[{\includegraphics[width=1.1in,height=1.35in,clip,keepaspectratio]{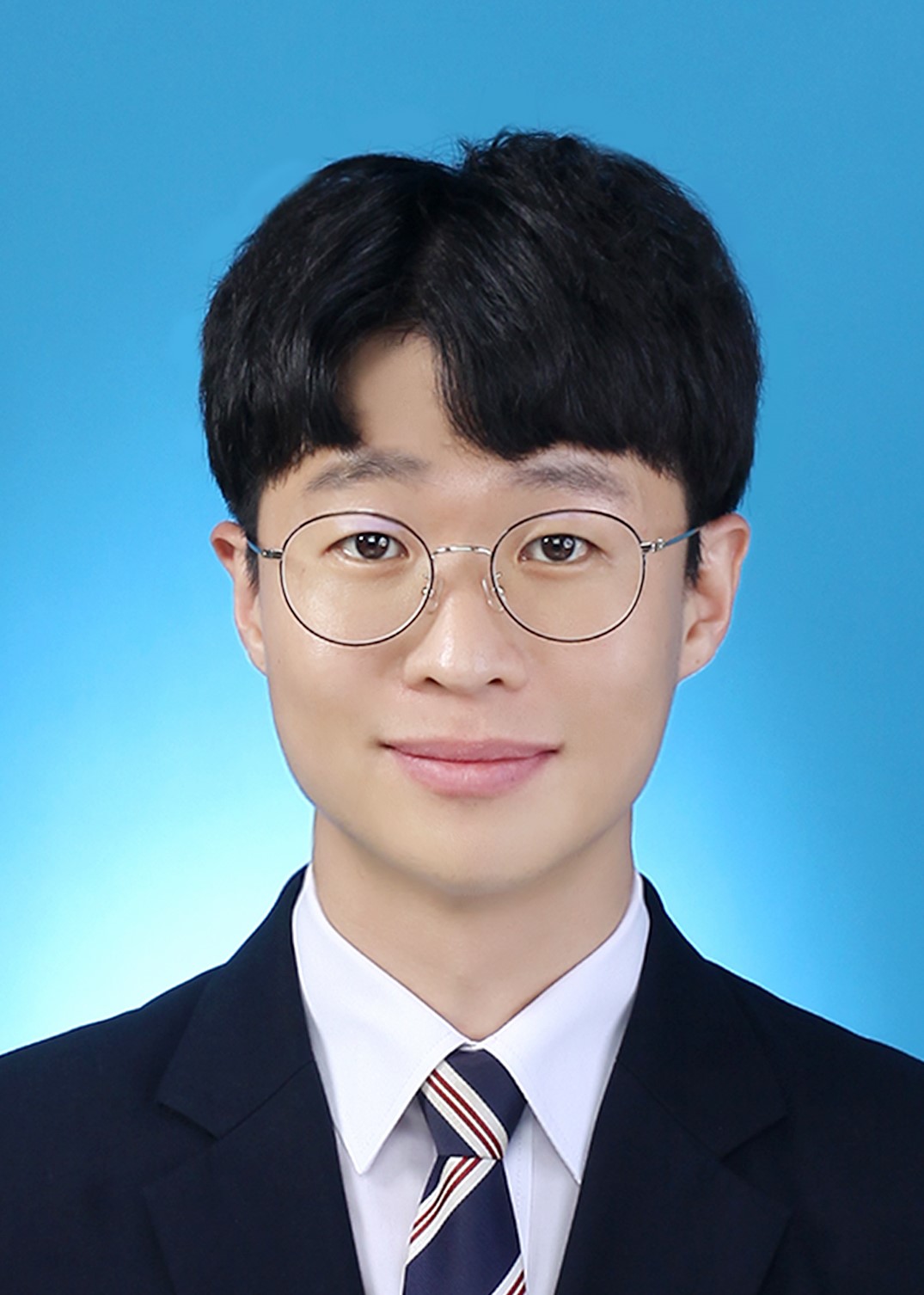}}]%
	{Hyoyoung Jung}
	(S'12) received the B.S. degree in electronic engineering from Inha University, South Korea, in 2011. He is currently pursuing the M.S. and Ph.D. integrated degree in electrical engineering and computer science from the Gwangju Institute of Science and Technology, South Korea. His research interests include statistical signal processing, machine learning, and anti-jamming communication systems.
\end{IEEEbiography}

\begin{IEEEbiography}[{\includegraphics[width=1.1in,height=1.35in,clip,keepaspectratio]{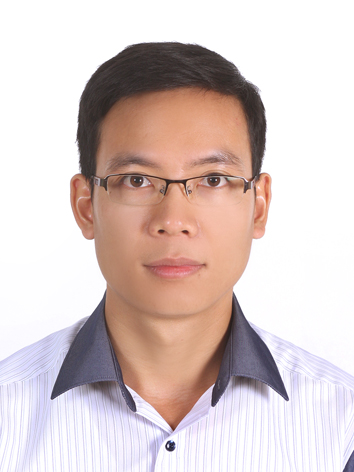}}]%
	{Binh Van Nguyen}
	received the bachelor degree in electrical and electronics from the Ho Chi Minh City University of Technology, Ho Chi Minh City, Vietnam, in 2010, and the M.S. and the Ph.D. degree in wireless communications from the Gwangju Institute of Science and Technology (GIST), Gwangju, Republic of Korea, in 2012 and 2016, respectively. From Sep. 2016 to Oct. 2018, he worked as a Research Fellow at GIST. From Nov. 2018 to Aug. 2019, he worked as an Assistant Research Professor at GIST. He is now with Samsung Electronics, Suwon, Republic of Korea, and with the Institute of Research and Development, Duy Tan University,  Da Nang 550000, Vietnam. His research interests include cooperative communications, physical layer security, and chaotic and anti-jamming communications.
\end{IEEEbiography}

\begin{IEEEbiography}[{\includegraphics[width=1.1in,height=1.35in,clip,keepaspectratio]{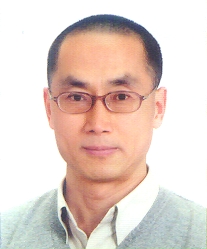}}]%
	{Iickho Song}
	(S’80–M’87–SM’96–F’09) received the B.S.E. \textit{(magna cum laude)} and M.S.E. degrees in electronics engineering from Seoul National University, Seoul, South Korea, in 1982 and 1984, respectively, and the M.S.E. and Ph.D. degrees in electrical engineering from the University of Pennsylvania, Philadelphia, PA, USA, in 1985 and 1987, respectively. 
	
	He was a Member of the Technical Staff, Bell Communications Research in 1987. In 1988, he joined the School of Electrical Engineering, Korea Advanced Institute of Science and Technology, Daejeon, South Korea, where he is currently a Professor. He has coauthored a few books including \textit{Advanced Theory of Signal Detection} (Springer, 2002) and \textit{Random Variables and Random Processes} (Freedom Academy, 2014; in Korean), and published papers on signal detection, statistical signal processing, and applied probability. 
	
	Prof. Song is a Fellow of the Korean Academy of Science and Technology (KAST). He is also a Fellow of the IET, and a Member of the Acoustical Society of Korea (ASK), Institute of Electronics Engineers of Korea (IEEK), Korean Institute of Communications and Information Sciences (KICS), Korea Institute of Information, Electronics, and Communication Technology, and Institute of Electronics, Information, and Communication Engineers. He has served as the Treasurer of the IEEE Korea Section, an Editor for the \textit{Journal of the ASK}, an Editor for \textit{Journal of the IEEK}, an Editor for the \textit{Journal of the KICS}, an Editor for the \textit{Journal of Communications and Networks (JCN)}, and a Division Editor for the \textit{JCN}. He was the recipient of several awards including the Young Scientists Award (KAST, 2000), Achievement Award (IET, 2006), and Hae Dong Information and Communications Academic Award (KICS, 2006).
\end{IEEEbiography}

\begin{IEEEbiography}[{\includegraphics[width=1.1in,height=1.35in,clip,keepaspectratio]{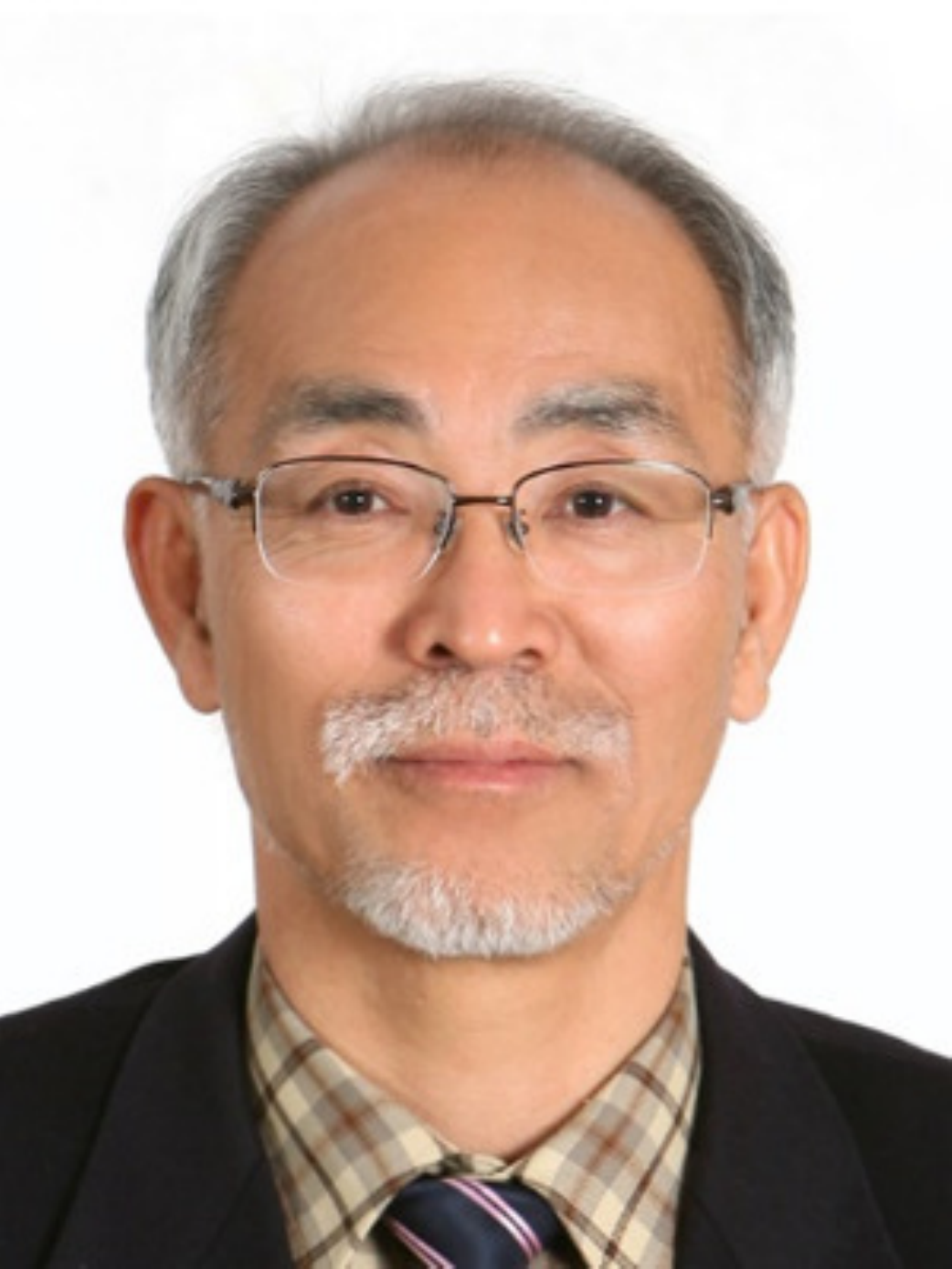}}]%
	{Kiseon Kim}
	received the B.Eng. and M.Eng. degrees in electronics engineering from Seoul National University, Seoul, South Korea, in 1978 and 1980, respectively, and the Ph.D. degree in electrical engineering systems from the University of Southern California at Los Angeles, Los Angeles, CA, USA, in 1987. From 1988 to 1991, he was with Schlumberger, Houston, TX, USA. From 1991 to 1994, he was with the Superconducting Super Collider Laboratory, Waxahachie, TX, USA. In 1994, he joined the Gwangju Institute of Science and Technology, Gwangju, South Korea, where he is currently a Professor. His current research interests include wideband digital communications system design, sensor network design, analysis and implementation both at the physical and at the resource management layer, and biomedical application design. He is also a member of the National Academy of Engineering of Korea, a Fellow of the IET, and a Senior Editor of the \textit{IEEE SENSORS JOURNAL}.
\end{IEEEbiography}
\end{document}

%% file: ref-jhy-p1-is-21.tex

%
